\documentclass[12pt]{article} 
\usepackage{amssymb}
\renewcommand{\Bbb}{\bf}
\newcommand{\newsubsection}[1]{
\vspace{8mm}
\pagebreak[3]
\addtocounter{subsection}{1}
\addcontentsline{toc}{subsection}{\protect
\numberline{\arabic{section}.\arabic{subsection}}{#1}}
\noindent{\it \thesubsection.  #1}                 
\nopagebreak
\vspace{2mm}
\nopagebreak}
\newcommand{\newsection}[1]{
\vspace{15mm}
\pagebreak[3]
\addtocounter{section}{1}
\setcounter{equation}{0}
\setcounter{subsection}{0}
\setcounter{footnote}{0}
\addcontentsline{toc}{section}{\protect
\numberline{\arabic{section}}{{\rm #1}}}
\noindent{\bf \thesection.  #1}                 
\nopagebreak
\vspace{4mm}
\nopagebreak}
\renewcommand{\theequation}{\thesection.\arabic{equation}}

\newlength{\extraspace}
\newlength{\extraspaces}
\newcounter{dummy}
\newcommand{\be}{\begin{equation}
\addtolength{\abovedisplayskip}{\extraspaces}
\addtolength{\belowdisplayskip}{\extraspaces}
\addtolength{\abovedisplayshortskip}{\extraspace}
\addtolength{\belowdisplayshortskip}{\extraspace}}
\newcommand{\ee}{\end{equation}}
\newcommand{\ba}{\begin{eqnarray}
\addtolength{\abovedisplayskip}{\extraspaces}
\addtolength{\belowdisplayskip}{\extraspaces}
\addtolength{\abovedisplayshortskip}{\extraspace}
\addtolength{\belowdisplayshortskip}{\extraspace}}
\newcommand{\ea}{\end{eqnarray}}
\newcommand{\nonu}{\nonumber \\[2mm]}
\newcommand{\is}{& \!\! = \!\! &}
\newcommand{\ban}{\begin{eqnarray*}
\addtolength{\abovedisplayskip}{\extraspaces}
\addtolength{\belowdisplayskip}{\extraspaces}
\addtolength{\abovedisplayshortskip}{\extraspace}
\addtolength{\belowdisplayshortskip}{\extraspace}}
\newcommand{\ean}{\end{eqnarray*}}
\newcommand{\baa}{                         
\addtocounter{equation}{1}
\setcounter{dummy}{\value{equation}}
\setcounter{equation}{0}
\renewcommand{\theequation}{\thesection.\arabic{dummy}\alph{equation}}
\begin{eqnarray}
\addtolength{\abovedisplayskip}{\extraspaces}
\addtolength{\belowdisplayskip}{\extraspaces}
\addtolength{\abovedisplayshortskip}{\extraspace}
\addtolength{\belowdisplayshortskip}{\extraspace}}
\newcommand{\eaa}{                                       
\end{eqnarray}
\setcounter{equation}{\value{dummy}}
\renewcommand{\theequation}{\thesection.\arabic{equation}}}

\input epsf
\newcounter{fignum}
\newcounter{tabnum}
\newcounter{xxx}
\newcommand{\bl}{\begin{list}{({\it\roman{xxx}})}{\usecounter{xxx}}}
\newcommand{\el}{\end{list}}
\renewcommand{\d}{{{\partial}}}

\newcommand{\ppt}[1]{{\partial \over \partial t}}            
\newcommand{\ppx}[1]{{\partial \over \partial x}}            
\newcommand{\pqt}[1]{{\partial^2 \over \partial t^2}}            
\newcommand{\pqx}[1]{{\partial^2  \over \partial x^2}}            
\newcommand{\ie}{{\it i.e.}}

\newcommand{\eg}{{\it e.g.\ }}

\renewcommand{\.}{\cdot}

\newcommand{\half}{{\textstyle{1\over 2}}}

\newcommand{\Z}{{\Bbb Z}}
\newcommand{\R}{{\Bbb R}}

\newcommand{\cH}{{\cal H }}

\newcommand{\ra}{\rightarrow}

\newcommand{\mmod}[1]{\ (\mbox{mod}\;#1)}

\newcommand{\lra}{{\mathop{\leftrightarrow}}}

\newcommand{\inv}{^{-1}}

\newcommand{\Tr}{{\rm Tr}\,}

\newcommand{\Lbar}{{\overline{L}}}

\newcommand{\cN}{{\cal N}}

\newcommand{\option}[4]{ \left\{ \begin{array}{ll}
#1, & \hbox{#2}, \\[2mm]
#3, & \hbox{#4}. \end{array} \right. }

\newcommand{\ext}{{\raisebox{.2ex}{$\textstyle \bigwedge$}}}

\def\e{\epsilon}

\def\F{\Phi}

\def\<{\langle}
\def\>{\rangle}

\newfont{\gothic}{eufm10 scaled\magstep1}

\begin{document}

\begin{center}

\addtolength{\baselineskip}{.5mm}
\thispagestyle{empty}
\begin{flushright}
December 1999\\
{\sc hep-th/9912101}\\
\end{flushright}

\vspace{34mm}

{\Large \sc Discrete Torsion and Symmetric Products}
\\[35mm]
{\sc Robbert Dijkgraaf}\\[5mm]
{\it Departments of Mathematics and Physics\\
University of Amsterdam\\
Plantage Muidergracht 24, 1018 TV Amsterdam}\\[3mm]
{\tt rhd@wins.uva.nl}\\[2cm] 
{\sc Abstract}
\end{center}

\begin{quote} 

\noindent In this note we point out that a symmetric product orbifold
CFT can be twisted by a unique nontrivial two-cocycle of the
permutation group. This discrete torsion changes the spins and
statistics of corresponding second-quantized string theory making it
essentially ``supersymmetric.'' The long strings of even length become
fermionic (or ghosts), those of odd length bosonic. The partition
function and elliptic genus can be described by a sum over stringy
spin structures. The usual cubic interaction vertex is odd and
nilpotent, so this construction gives rise to a DLCQ string theory
with a leading quartic interaction.

\end{quote}

\vfill

\newpage

\newsection{Introduction}

Symmetric product orbifolds are two-dimensional
sigma models on the configuration space
\be 
S^N\! X = X^N/S_N 
\ee 
of $N$ unordened points on a space $X$. Although this is a singular
space, since the permutation group $S_N$ does not act free, the
corresponding CFT is well-behaved. In the last years these CFT's have
played an important role in string theory.  Symmetric products appear
naturally as moduli spaces of supersymmetric gauge theories and
world-volume theories of D-branes, and the corresponding CFT's are a
crucial ingredient in matrix string theory
\cite{motl,banks-seiberg,dvv} and the light-cone gauge quantization
of so-called little string theories
\cite{dvv2,witten,abkss}. (See also \cite{sym} for background material on
symmetric product models.) In all these applications one describes
second-quantized string theories by considering a single sigma model
with target $S^N\!X$, effectively second-quantizing the spacetime
first.

In this note we want to point out that these orbifold conformal field
theories have a natural simple generalization. One can include a nontrivial
discrete torsion class
\be
H^2(S_N,U(1)) = \Z_2
\label{cocycle}
\ee
for the action of the permutation group. With this torsion the
configuration space becomes in some mild sense a non-commutative
space.  The discrete torsion class defines a spin cover of the
permutation group where disjoint elementary transpositions anticommute
instead of commute. We will point out that physically the string
theory becomes essentially supersymmetric, with equal number of
bosonic and fermionic strings, where the statistics are determined by
the winding number of the string (with even windings fermionic and odd
winding bosonic).

The plan of this note is as follows. We will first recall the
definition and some interpretations of discrete torsion in orbifold
conformal field theories. Then we will apply this formalism to the
symmetric group in section 3, where we explicitely compute the effects
of the two-cocycle (\ref{cocycle}). We then turn to the application in
CFT and second-quantized string theory in sections 4 and 5. The
partition function, and in particular the elliptic genus, is described
in section 6. We end with some concluding remarks and speculations.

\newsection{Discrete torsion}

Consider an orbifold conformal field theory that is obtained by
quotienting by a finite group $G$. Then it is well-known that for any
non-trivial two-cocycle 
\be
c\in H^2(G,U(1))
\ee
in the group cohomology of $G$ we can define a new model by weighting
the twisted sectors of the orbifold with a non-trivial phase, the
so-called discrete torsion \cite{vafa}. In the case of a sigma model on
a geometric orbifold $X/G$ this discrete torsion can be seen as a
particular choice of a flat $B$-field --- or ``flat gerbe'' in modern
parlance \cite{gerbe} --- on the (possibly singular) target space.

One way to understand discrete torsion is to realize that orbifold
quantum field theories
can be considered as discrete gauge theories with gauge group $G$. 
In a Lagrangian
formulation the partition function of such an orbifold theory on a Riemann
surface $\Sigma$ is obtained by summing over all possible $G$ bundles
over $\Sigma$.  Topologically these bundles are classified by homotopy
classes of maps of $\Sigma$ into the classifying space $BG$, the base
space of the universal $G$ bundle. We can identify the group
cohomology $H^*(G)$ with the usual cohomology of the classifying space $BG$.  
In the path-integral we will sum over all maps
\be
x\colon\ \Sigma \to BG,
\ee
and in the presence of a discrete torsion class $c\in H^2(BG,U(1))$ we 
weight the map $x$ by an extra phase factor \cite{dw}
\be
\langle x^*c,\Sigma\rangle \in U(1).
\ee

Alternatively, in a Hamiltonian formalism the Hilbert space $\cH$ of 
the orbifold decomposes into sectors as
\be
\cH = \bigoplus_{[g]} \cH_g
\ee
Here the superselection sectors $\cH_g$ consist of states 
twisted by $g\in G$ and are labeled by the conjugacy class $[g]$. 
The twisted states still carry a residual action of the centralizer
$C_g$, the subgroup consisting of all elements of $G$ that commute with
the twist $g$. Normally, in an orbifold theory one projects in the twisted 
sector $\cH_g$ on the states invariant under $C_g$. But,
in the presence of discrete torsion the projection picks out the
states that transform as a particular (possibly
trivial) one-dimensional representation of $C_g$. 
This representation of $C_g$, that we will write as $\e(g,\.)$,
is given in terms of the group 2-cocycle $c(g,h)$ as
\be
\e(g,h) = {c(g,h) \over c(h,g)},\qquad hg=gh.
\ee
One can think of the representation $\e(g,h)$ geometrically as the phase factor 
that is associated to the (flat) $G$ bundle on the two-torus $T^2$ given by
the commuting pair of holonomies $(g,h)$. By this geometric
interpretation, or equivalently by the cocycle condition on $c$, the
above definition of $\e(g,h)$ is manifest $SL(2,\Z)$ invariant. 

One concrete way to think about this particular 
one-dimensional representation of $C_g$ is
as follows. A two-cocycle or ``Schur multiplier'' $c\in H^2(G,U(1))$
defines a central extension $\widehat G$ of $G$,
\be
1 \to U(1) \ra \widehat G \ra G \to 1.
\ee
If $G$ is a finite group, then  
\be
Z := H^2(G,U(1)) \cong H^3(G,\Z)
\ee
is a finite abelian group and the central extension can be restricted to a
finite extension by $Z$,
\be
1 \to Z \ra \widehat G \ra G \to 1.
\ee
Given such a central extension, and a pair of commuting elements $g,h$
in the group $G$, we can lift the pair to elements $\hat g,\hat h$ in
the covering group $\widehat G$. Now in this central extension the lifts do
no longer necessarily commute, and the commutator of the elements
$\hat g$ and $\hat h$ gives the required one-dimensional representation
that takes values in $Z$. That is, we have
\be
\e(g,h) = [\hat g,\hat h] = \hat g \,\hat h\, \hat g\inv \hat h\inv \in Z.
\ee
Note that the commutator is independent of the choices of lifts, since
these choices differ, by definition, by a central element, and these
central elements cancel out in the commutator.

\newsection{Discrete torsion for the symmetric group}

Let us now consider the case where the finite group $G$ is taken to be
the symmetric group on $N$ elements $S_N$. In this case it is
well-known that in the ``stable range''  $N\geq 4$
\be
H^2(S_N,U(1)) \cong \Z_2.
\ee
That is, there is a unique non-trivial central extension of the
permutation group
\be
1 \to \Z_2 \to \widehat S_N \to S_N \to 1.
\ee
(In the case $N<4$ this extension by $\Z_2$ still exists, but it is
trivial when considered as an extension by $U(1)$. For example for
$N=2$ we get the familiar extension $\Z_2 \to \Z_4 \to \Z_2$.)

Geometrically we can think of this double cover as follows. Consider the
natural action of $S_N$ on an orthonormal basis of the vector space
$\R^N$. This gives an embedding of $S_N$ into $O(N-1)$, the orthogonal
group that acts on the hyperplane that contains the $N$ basis vectors.
The orthogonal group has a unique $\Z_2$ central extension $Pin(N-1)$.
Restricted to the rotation group $SO(N-1)$ it is the usual spin cover
$Spin(N-1)$. The restriction of this spin cover to the symmetric group
defines the $\Z_2$ central extension that we are looking for.

It should be noted that in parastatistics, where one considers
higher-dimensional representations of the permutation group, this
central extension plays an important role, since it leads to spinor
representation of the statistics group, see \eg\ \cite{wilczek}.
Such representations where first studied by Schur, and play an
important role in the statistics of quasiparticles in the Pfaffian
$\nu=\half$ quantum Hall state.

In terms of generators and relations we can define the spin cover
$\widehat S_N$ as follows \cite{hoffman-humphreys,karpilovsky}. Let us
denote the standard generators of $S_N$ by $t_i$ with $i=1,\ldots,N-1$.
Here $t_i$ is the elementary transposition that interchanges the
elements $i$ and $i+1$. These generators satisfy the familiar relations
\ba
t_i^2 \is 1,\nonu
t_it_{i+1} t_i \is t_{i+1} t_i t_{i+1},\\[2mm]
t_i t_j \is  t_j t_i,\qquad j>i+1.\nonumber
\ea
The central extension is obtained by replacing these generators by
their lifts $\hat t_i$, adjoining the central element $z$ with $z^2 =
1$ (we will often write $z=-1$) and modifying the above relations to
\ba
\hat t_i^2 \is z,\nonu
\hat t_i \hat t_{i+1} \hat t_i \is \hat t_{i+1} \hat t_i \hat
t_{i+1},\\[2mm]
\hat t_i \hat t_j \is  z \, \hat t_j \hat t_i,\qquad j>i+1. \nonumber
\ea

Note that the lift of a transposition has order four, and that two
transpositions that act disjunctly anticommute. That is because in the
spin extension these operations, which are geometrically (Weyl)
reflections, are represented by Dirac matrices and therefore
anticommute instead of commute. Indeed this last fact is responsible
for the non-trivial two-cocycle of $S_N$. It comes from the $\Z_2
\times \Z_2$ subgroup that is generated by interchanging particles 1
and 2 and particles 3 and 4. These generators now satisfy
\be
\hat t_1 \cdot \hat t_3 = - \hat t_3 \cdot \hat t_1.
\ee
The discrete torsion of the symmetric group is therefore simply the lift of 
the group cocycle (so familiar to string theorists) that generates
\be
H^2(\Z_2 \times \Z_2,U(1)) \cong \Z_2.
\ee
This also explain why the discrete torsion only appears for $N \geq 4$. 

Using these generators and relations it is now completely straightforward to
compute explicitly the phases $\e(g,h)$ in the twisted sectors. 
Here we should keep in mind that any element $g\in S_N$ is conjugate
to a product of cyclic permutations, that can be labeled by a partition
of $N$
\be
[g] = (1)^{N_1} (2)^{N_2} \cdots,\qquad \sum_{n\geq 1} n N_n =N.
\ee
The centralizer of such an element $g$ is given by
\be
C_g \cong \prod_{n\geq 1} S_{N_n} \ltimes \Z^{N_n}_n
\ee
Since $\e(g,h)$ is a representation of $C_g$, it satisfies
\be
\e(g,h_1h_2) = \e(g,h_1) \e(g,h_2).
\ee
Therefore it suffices to compute the phases $\e(g,h)$ for the two
specific kinds of elements $h$ that together generate $C_g$: 

\medskip

\noindent (1) a generator of the cyclic group $\Z_n$ of order $n$;

\noindent (2) an elementary transposition in $S_{N_n}$ that permutes two of
those cycles of length $n$. 

\medskip

We will see that the analysis (and the answer) depends critically on
the overall signature or parity $|g| = 0,1$ (mod 2) of the element
$g$, that we note can be written as
\be 
|g| = \sum_{n\geq 1} (n-1) N_n \mmod 2.
\ee
For example, since we have the fundamental result that for $|i-j|>1$
\be
\e(t_i,t_j) = [\hat t_i, \hat t_j] = -1,
\ee 
we will often use the fact that, if two elements $g$ and $h$ act 
disjunctly, they either commute or they anticommute in the extension
$\widehat S_N$, depending on their signs,
\be
\e(g,h)= (-1)^{|g||h|}.
\ee

\medskip

Case (1). 
Let us first consider a sector twisted by an element $g$ that contains a
cycle of length $n$. Denote the generator of that cycle $k$, with
$k^n=e$. So we can write $g = k \. g'$ where $g'$ commutes with $k$.
Since $g'$ and $k$ act disjunctly we can conclude that
\be
\e(g',k)= (-1)^{|g'||k|},
\ee
and, because $\e(g,k)=\e(g',k)$, we therefore find
\be
\e(g,k) = \option {1}{$n$ odd} {(-1)^{|g|-1}}{$n$ even}
\label{proj}
\ee
We will give an interpretation of this result later.

\medskip

Case (2). Let us now consider a permutation $x_n$ that interchanges two
disjoint cycles of equal length $n\geq 2$ in $g$. We denote the generators of
these two cyclic group as $k'$ and $k''$ and write $k=k'\. k''$. We
first compute the phase $\e(x_n,k)$. The only non-trivial computation is
actually the case $n=2$, since we will see that the general case follows
directly from this. 

In the case $n=2$ we can choose $k'=t_1$ and $k''=t_3$. The exchange 
operator $x_2$ satisfies by definition 
\be
x_2\. t_1 = t_3 \. x_2,
\ee
and can be written as
\be
x_2 = t_2 t_1 t_3 t_2.
\ee
With some simple algebra one now finds with $k=t_1t_3$
\ba
\hat k \. \hat x_2 \is
\ \ \hat t_1 \hat t_3 \hat t_2 \hat t_1 \hat t_3 \hat t_2 =
-\hat t_3 \hat t_1 \hat t_2 \hat t_1 \hat t_3 \hat t_2 \nonu
\is -\hat t_3 \hat t_2 \hat t_1 \hat t_2 \hat t_3 \hat t_2 
=  - \hat t_3 \hat t_2 \hat t_1 \hat t_3 \hat t_2 \hat t_2 \nonu
\is \ \ \hat t_3 \hat t_2 \hat t_3 \hat t_1 \hat t_2 \hat t_2 
= \ \ \hat t_2 \hat t_3 \hat t_2 \hat t_1 \hat t_2 \hat t_2 \nonu
\is \ \  \hat t_2 \hat t_3 \hat t_1 \hat t_2 \hat t_1 \hat t_2 
= -\hat t_2 \hat t_1 \hat t_3 \hat t_2 \hat t_1 \hat t_2 =
- \hat x_2 \. \hat k,
\ea
so that 
\be
\e(x_2,k) = -1.
\ee
Since $g$ can be written as $g=g' \. k$, where $g'$ acts disjunctly
from $x_2$, and since $x_2$ has even parity, we find
\be
\e(x_2,g)=\e(x_2,k)= -1.
\ee
This result can be formulated as that with discrete
torsion two cycles of length 2 anticommute instead of commute.

In the general case, where we consider an exchange of two cycles of length
$n \geq 2$, we simply should observe that the element $x_n$ has
conjugacy class $(2)^n$. The element $k$ acts by a cyclic
permutation of length $n$ on these $n$ two-cycles. That is,
$k$ is a product of $n-1$ elements that each exchange a pair
of two-cycles and therefore are conjugated to $x_2$.
Therefore we find
\be
\e(x_n,g)  = \e(x_n,k) = (-1)^{n-1}.
\label{stat}
\ee
This has a straightforward interpretation. When we exchange two $n$-cycles
we obtain an extra minus sign if $n$ is even. So $n$-cycles
behave as ``bosons'' if $n$ is odd, and as ``fermions'' if $n$ is even.

\newsection{The CFT interpretation}

We will now apply the above computations to determine the effect of
discrete torsion on the symmetric product orbifold conformal field
theory.

\newsubsection{Symmetric product orbifolds}

Suppose we start with a conformal field theory $X$ with Hilbert space
$\cH=\cH(X)$. We want to determine the Hilbert space of the symmetric
orbifold $ S^N\!X=X^N/S_N$. In the case without discrete torsion the
answer has been determined in \cite{dmvv}. (See also
\cite{permutation} for details about the chiral structure such as
fusion rules, modular transformations and braiding matrices of
rational permutation orbifold CFT's.)

The answer is most elegantly formulated if one considers the direct sum
of all symmetric products $S^N\!X$ summed over $N$. Then the resulting
Hilbert space has the structure of a Fock space, generated by an
infinite set of Hilbert spaces $\cH_n$, $n\geq 1$, based on the target
space $X$. 

More precisely, if we use the formal generating space of symmetric
powers with dummy variable $p$
\be
S_p\cH = \sum_{N\geq 0} p^N S^N\cH,
\ee
(and similarly $\ext_p\cH$ for the generating space of exterior powers),
then we have the result
\be
\sum_{N\geq 0} p^N \cH(S^N\!X) = 
\bigotimes_{n>0} S_{p^n}\cH_n(X),
\ee
where the Hilbert spaces $\cH_n(X)$ are defined as the subspace of
$\cH(X)$ satisfying
\be
L_0 - \Lbar_0 = 0 \mmod n.
\label{level}
\ee
The Hilbert spaces $\cH_n$ carry redefined Hamiltonians
$L_0^{(n)}=L_0/n$. One refers to these states generally as ``long
strings'' \cite{maldacena-susskind} of length $n$.

Note that the sector $\cH_g$ of the orbifold $S^N\!X$ twisted by a
group element $g$ of conjugacy class
\be
[g] = \prod_{n\geq 1} (n)^{N_n}
\ee
corresponds to the summand 
\be
\bigotimes_{n\geq 1} S^{N_n}\cH_n
\ee
in the above expression. 

In all these formulas it should be remembered that our starting point
is an action of the symmetric group on $S^N\cH$. Here we have some
freedom in how we want $S_N$ to act. In general, the space $\cH$ will
be graded and the corresponding action of $S_N$ will respect this
gradation. That is, it will act in the appropriate way by
symmetrization or anti-symmetrization on the even or odd parts. For
example if $\cH$ is the Hilbert space of a free fermion, it will be
$\Z_2$ graded by the fermion number. All actions of the permutation
group are always assumed to be in this graded sense.

\bigskip

\newsubsection{Symmetric products with torsion}

Let us now see how these formulas are changed if one includes discrete
torsion. First of all we have to distinguish between twisted sectors
related to $n$-cycles with $n$ odd and even. According to our formula
(\ref{stat}) the statistics of these sectors is now commuting or
anticommuting depending on whether $n$ is odd or even. We will refer
to these states as bosons and fermions. The ``fermion number'' $F$
(mod 2) of a twisted state simply equals the sign of the corresponding
cyclic permutation. The total parity $|g|$ of the twisted sector $g$
thus reflects the total fermion number of the state. Note that this
combinatorial fermion number should be added to the fermion number
that might already be present in the gradation of the one-string
Hilbert space $\cH(X)$.

Secondly, for a sector corresponding to a $n$-cycle we also have to
implement the $\Z_n$ projection. As we have seen, in the case without
discrete torsion invariance under the action of this $\Z_n$ subgroup
changes the conventional level matching condition. Instead of the
requirement that the spin $L_0-\Lbar_0$ is integer, one now restricts to
the modified condition that in the Hilbert space $\cH_n$ the spin is an
$n$-fold as in (\ref{level}). 

With discrete torsion turned on, formula (\ref{proj}) tells us that we
have to project differently in the case that $n$ is even and $(-1)^{|g|}
= 1$. In that case we no longer require invariance under $\Z_n$ but
instead only keep the states in the sign representation where the generator 
of $\Z_n$ is represented as $-1$.
So the new level matching condition gives the following quantization for
the spin $m$ of strings of even length $n$:
\be
m := L_0^{(n)} - \Lbar_0^{(n)} = {1\over n}(L_0 - \Lbar_0) \in \Z + \half.
\ee
That is, in the orbifold CFT the ``fermionic'' states of
even length $n$ will have half-integer spin. Since $L_0-\Lbar_0$ is
always an integer, this condition clearly only makes sense for $n$ even.
Note that the overall spin of this state is still integer,
because we only apply this construction in the case that the total
fermion number is even.
We will denote quite generally as $\cH_n^{\pm}$ the subspaces
of $\cH_n$ of (half)integer spin, \ie, the subspaces consisting
of states that satisfy $(-1)^{2m}=\pm 1$.

Adding everything together the Hilbert space of the symmetric
product orbifold $S^N\!X$ with the nontrivial 2-cocycle $c\in
H^2(S_N,U(1))$ takes the following form
\ba
\cH^c(S^N\!X) \is \bigoplus_{\scriptstyle  even\, \{N_n\} \atop 
\scriptstyle \sum nN_n = N\ } \bigotimes_{n>0}
S_{p^{2n-1}}\cH^+_{2n-1} \otimes \ext_{p^{2n}} \cH^-_{2n} \nonu
&&\qquad
\bigoplus_{\scriptstyle odd\, \{N_n\} \atop 
\scriptstyle \sum nN_n = N\ } \bigotimes_{n>0}
S_{p^{2n-1}}\cH^+_{2n-1} \otimes \ext_{p^{2n}} \cH^+_{2n}
\ea
Here a partition $\{N_n\}$ is called even or odd
depending on the parity $|g|$ of the corresponding permutation.

\newsection{Second-quantized strings}

Let us make some remarks on the physical interpretation of this result
in string theory. Of course the most obvious effect is that the
discrete torsion has changed the statistics of the strings. The ``long
strings'' of even length have turned into fermions, while the strings
of odd length are still bosons. Since we roughly have as many bosons
as fermions the model looks ``supersymmetric,'' although, as we will
point out, it is perhaps better to refer to ghosts instead of
fermions. The second, more subtle effect of discrete torsion is its
influence on the quantization of the spins of the fermionic
strings. This interpretation is made more precise in terms of the
light-cone quantization formalism.

\newsubsection{Interpretation in DLCQ}

Recall that permutation orbifold CFTs on the configuration space
$S^N\!X$ appear naturally in the description of second-quantized
strings theories on
\be
X \times \R^{1,1}
\ee
in light-cone gauge, where the longtudinal momentum $p^+$ is
discretized, the so-called discrete light-come quantization or
DLCQ. In this quantization scheme the null coordinate $x^+$ is used to
describe the time evolution of the system, whereas the null coordinate
$x^-$, that is canonically conjugated to $p^+$, is assumed to be
periodic. This interpretation of the symmetric product orbifold is
used in the matrix string interpretation of the ten-dimensional Type
IIA string and in the six-dimensional little string theories related
to coinciding fivebranes. (See also \cite{little} for a review of
little string theories.)

In this interpretation the quantum number $n$, the length of the string,
is proportional to the discrete momentum $p^+$. In fact, for our
purposes it is best to identify it as
\be
p^+ = {n\over 2}
\ee
We want to think of the states with $p^+$ integer or half-integer
as fields that are periodic or anti-periodic around the compact
direction $x^-$. This is a well-known setup for the best-known example
of a DLCQ field theory: a chiral conformal
field theory in $1+1$ dimensions. Here the momentum $p^+$ is simply the
eigenvalue of the operator $L_0$ (not to be confused with the
world-sheet $L_0$ used above) and we know that for fermionic field
this eigenvalue will be half-integer in the NS spin structure
and integer in the R spin structure.

However, something strange is going on here. According to our
computation it is the states with $p^+$ integer that have turned into
fermions, whereas the bosons have half-integer $p^+$ respectively. Since
it is the bosons that are sensitive to the spin structure and get
naturally anti-periodic boundary conditions, we see that from this point
of view we are actually dealing with fields that are more appropriately
called ghosts, since the usual relation between spin and statistics is
not satisfied. Because of the clash with the spin-statistics theorem two
bosons can decay into a fermion {\it etc.}

To understand the second effect of discrete torsion
we should remember that strings in DLCQ
carry two natural quantum numbers: the longitudinal momentum $p^+=n/2$
discussed above together with a longitudinal winding number
$w^-$, an integer that measures the number of times a string is wrapped
around the compactified $x^-$ direction. In light-cone gauge the
coordinate $x^-$ is expressed in terms of the world-sheet stress-tensor
as
\be
\d x^- = {1\over p^+} T(z).
\ee
Therefore the winding number is given by
\be
w^- = {1 \over 2\pi} \oint dx^- = {1\over p^+}(L_0 - \Lbar_0)= 2m,
\ee
with $m$ the conformal field theory spin.
We have seen that $m$ can be half-integer,
so $w^-$ can be even or odd. But the combination
\be
p^+ w^- = L_0 - \Lbar_0 = nm
\ee
is always integer, as is required by level-matching. 

So in the presence of discrete torsion the fermionic strings, with $n$
even, can have different excitations depending on whether the total
fermion number is even or odd. If the total fermion number is even,
these fermionic strings have half-integer spin excitations, or
equivalently odd winding numbers. If the total fermion number is odd,
all winding numbers are even, as is always the case for the bosonic
strings.

This interpretation is still not very satisfying nor intuitive, so we
will now turn to the dual picture obtained by performing a T-duality.

\newsubsection{T-duality and stringy spin structures}

There is an interesting point concerning T-duality in the null
direction $x^-$. Such a duality will interchange the momentum $p^+$
and the winding number $w^-$, or equivalently the quantum numbers $n$
and $m$.  In the case without torsion this is a manifest symmetry of
the second-quantized theory. (In order to make this a full symmetry
one has to add the $p^+=0$ sector, which is always problematic in
light-cone quantization. See also the discussion in \cite{dyon}.) This
symmetry between momentum and winding modes seems to be broken in the
presence of the discrete torsion, since it is momentum $p^+$ that
determines the statistics of the strings.

In fact, the complete physical interpretation becomes more conventional
if we work in this T-dual framework. If we interchange (to be precise)
$p^+$ and $w^-/2$, we have
\be
w^-=n,\qquad p^+=m,
\ee
and strings that are wrapped an even number of times are fermions, and
those that are wrapped an odd number of times are bosons. It would be
obvious to assume that unwrapped strings with $n=0$ are also
fermionic, although they do not appear explicitly in this DLCQ scheme.

Since we have fermionic fields, one can think of making a spin
projection of the model, by summing over all spin structures along the
$x^-$ direction and projecting on even fermion number --- a procedure
implemented by summing over spin structures in the time direction
$x^+$.  This would be the spacetime equivalent of the familiar
procedure in two dimensional CFT, where the sum over spin structures
produces out of a fermionic non-local model a bosonic local
model. However, introducing spacetime spin structures in string theory
introduces an extra complication, because fermions can also carry
winding numbers.

Indeed, consider the general situation of a string on a spacetime that
contains a $S^1$, say for convenience of radius one.\footnote{See also
the closely related discussion in \cite{vector}.} On this circle we have
momenta $p\in \Z$ and winding numbers $w\in \Z$. Level-matching of the
CFT will always require that
\be
p\cdot w \in \Z.
\ee
Consider now a (spacetime) fermion mode of the string. We have to pick a
spin structure on the circle that determines the boundary conditions for
such a field. There are two choices: the anti-periodic (Neveu-Schwarz)
spin structure that gives $p\in \Z + \half$ and the periodic (Ramond)
spin structure that quantizes the momentum as $p\in \Z$. 

However, if the string also carries a winding number $w$ the story gets
a bit more complicated. For half-integer $p$ we cannot allow arbitrary
integer winding number, since level-matching requires $p\cdot w$ to be
integer. So we see that only string states with {\it even} winding
number can be fermions. This is the effect we have been observing in
this note: fermion statistics is only consistent for strings of even
length. Only they can naturally be anti-periodic. 

Now that we have seen that the even windings can couple to a spin
structure, it is naturally to sum over spin structures, both in the
compactified space direction (which might be null, as in DLCQ) as in the
time direction. The latter procedure has the familiar interpretation as
a projection on even fermion number. In the NS sector, where $p$ is
half-integer quantized and where there is a unique ground state, this
implies that the total parity $|g|$ should be even. This effect we have
seen. Only for $|g|$ even could we have half-integer $m$. In the R
sector the spin projection is ambiguous, because of the assignment of
fermion number to the ground state. According to our formulas we should
assign to the ground state odd fermion number, since we require that the
total parity $|g|$ is odd in this case.

\newsection{The elliptic genus}

It is interesting to translate these considerations into a concrete
formula for the genus one partition function for the orbifold theory
including the effect of the discrete torsion. This formula becomes
particularly simple if we restrict to the chiral partition function,
the so-called elliptic genus which for Calabi-Yau sigma models with
(at least) $\cN=2$ supersymmetry is defined as the following character
in the RR sector
\be
\chi(X;q,y) = \Tr_{\strut \! \cH_{RR}} \left[
(-1)^F y^{F_L} q^{L_0 - {c\over 24}}\right]
\ee
Given the Fourier expansion
\be
\chi(X;q,y)= \sum_{m,\ell} c(m,\ell) q^m y^\ell
\ee
of the chiral partition function of the sigma model on $X$, 
the generating function of the elliptic genera of the symmetric products is 
\be
Z(p,q,y) = \sum_{N \geq 0} p^N  \chi(S^N\!X;q,y) = \prod_{n>0,\,m,\ell}
(1-p^nq^m y^\ell)^{-c(nm,\ell)}
\ee

As we have sketched in the previous section, the partition function
(and therefore also the elliptic genus) for the orbifold with discrete
torsion is best written as a sum over spacetime spin structures
\be
Z(p,q,y) = {1\over 2} \left(Z_{++} + Z_{+-} + 
Z_{-+} + Z_{--}\right).
\ee
Here the contributions of the four spin structures are
\ba
Z_{++}(p,q,y) \is  \prod_{n>0,\, m} 
{\Bigl( 1 + p^{2n} q^{m - {1\over 2}}y^\ell\Bigr)^{c(n(2m-1),\ell)} \hspace{-17mm} 
\over 
\Bigl( 1 - p^{2n-1} q^{m}y^\ell\Bigr)^{c((2n-1)m,\ell)}\hspace{-17mm}} \nonu
Z_{+-}(p,q,y) \is  \prod_{n>0,\, m} 
{\Bigl( 1 - p^{2n} q^{m - {1\over 2}}y^\ell\Bigr)^{c(n(2m-1),\ell)} \hspace{-17mm} 
\over 
\Bigl( 1 - p^{2n-1} q^{m}y^\ell\Bigr)^{c((2n-1)m,\ell)}\hspace{-17mm}} \nonu
Z_{-+}(p,q,y) \is  \prod_{n>0,\, m} 
{\Bigl( 1 + p^{2n} q^{m}y^\ell\Bigr)^{c(2nm,\ell)} \hspace{-12mm}
\over 
\Bigl( 1 - p^{2n-1} q^{m}y^\ell\Bigr)^{c((2n-1)m,\ell)}\hspace{-17mm}} \nonu
Z_{--}(p,q,y) \is - \prod_{n>0,\, m} 
{\Bigl( 1 - p^{2n} q^{m}y^\ell\Bigr)^{c(2nm,\ell)} \hspace{-12mm}
\over 
\Bigl( 1 - p^{2n-1} q^{m}y^\ell\Bigr)^{c((2n-1)m,\ell)}\hspace{-17mm}} 
\ea

In the case with zero discrete torsion, after a so-called automorphic
correction that adds in the $p^+=0$ sector and the shift in the ground
state energy, the partition function takes the form
\be
\F(p,q,y)= (pq)^{-\chi/24} \prod_{n,m,\ell \geq 0} (1-p^nq^my^\ell)^{-c(nm,\ell)}.
\ee
The infinite product $\F$ will be typically an automorphic form of the
T-duality group $SO(2,3,\Z) \cong Sp(4,\Z)$ 
\cite{harvey-moore,kawai,dyon,neumann,gritsenko}. This is an example of the 
famous lifting of a modular form to an automorphic product as
discussed by Borcherds \cite{borcherds}.

It would be interesting to investigate the automorphic properties of the
infinite products that are obtained by including the effect of discrete 
torsion. Only for a given spin structure the partition
function can be computed as a one loop amplitude with target space
\be
X \times T^2
\ee
as in \cite{harvey-moore}. So automorphicity, which is simply the
T-duality associated to the light-cone torus $T^2$, is not as
straightforward to check. Notice in this respect that the proper
T-duality that interchanges $p$ and $q$ seems to be differently
realized in this fermionic model. For example, the partition function
$Z_{+-}$ transforms under the transformation $p\ \lra\ q^{1\over 2}$
as
\be 
\log Z_{+-}(q^{1\over 2},p^2,y) = - \log Z_{+-}(p,q,y)
\ee
which suggests that we have some nontrivial multiplier for the free
energy $\log Z$. It would also be interesting to know if these
second-quantized elliptic genera are naturally related to characters
of super-Lie algebras.

If we put $y=1$ only ground states states with $L_0=c/24$ contribute,
and the elliptic genus degerates to the Euler number or
Witten index.  For the symmetric product this gives the well-known
identity \cite{goettsche,hirzebruch,vafa-witten,sym}
\be
\sum_{N \geq 0} p^N  \chi(S^N\!X) =
 \prod_{n>0} (1-p^n)^{-c}
\ee
with $c=\chi(X)$ the Euler number of $X$. This is almost a modular
form for the $SL(2,\Z)$ action on $\sigma$, with $p=e^{2\pi i
\sigma}$.
With discrete torsion only the configurations with no or an odd number
of strings of even length contribute, so we get instead the expression
\be
\prod_{n>0} (1 - p^{2n-1})^{-c}\left[ 
1 + \half \prod_{n>0} (1 + p^{2n})^c -  \half \prod_{n>0} (1 - p^{2n})^c
\right]
\ee
No obvious modular properties remain.

\newsection{Concluding remarks}

Matrix string theory can be seen as a DLCQ version of string field
theory. It describes perturbative string theory by conformal
perturbation theory around the orbifold CFT $S^N\R^8$, where the
leading irrelevant operator can be identified with the Mandelstam
cubic string vertex \cite{dvv}. If we twist the CFT with our discrete
torsion we obtain a model with exotic statistics. In the large $N$
limit the momentum $p^+=n/2$ is send to infinity, keeping the ratio
$p^+/N$ finite and making it a continuous variable. The distinction
between even and odd $n$ disappears, and one gets truly equal number
of bosons and fermions.  

However, a cubic string vertex, which is realized as a $\Z_2$ twist
field, now has fermionic statistics. That is, the standard string
coupling constant $g_s$ becomes a {\it nilpotent} Grassmann variable
that squares to zero
\be
g_s^2=0,
\ee
a rather mystifying phenomenon. Because this cubic vertex is itself
fermionic, it mediates interactions where two fermionic strings (say
of length 2) can combine to another fermionic string (of length
4). 

The next to leading order perturbation is a $\Z_3$ twist
field. This represents in string perturbation a quartic contact
term. Its couping constant (which in the conventional setup is
proportional to $g_s^2$) remains bosonic and is now the leading
irrelevant deformation.

\medskip

The inclusion of discrete torsion is also interesting in the case of
the so-called D1-D5 system in Type IIB string theory compactified on a
four-torus or $K3$ manifold $X$. Such a D1-D5-brane represents a
string in the remaining six uncompactified dimensions. The infrared
limit of the world-sheet theory gives rise to a $\cN=(4,4)$ SCFT on
the moduli space of instantons on $X$. The central charge is given by
$c=6k$ with $k=Q_1Q_5$, the product of the number of D1-branes and
D5-branes. The number of real marginal deformations of this SCFT is
$4\times 5 $ or $4 \times 21$ for $X=T^4$ or $K3$ respectively.

For certain values of the space-time moduli this hyperk\"ahler moduli
space coincides with the symmetric product $S^N\!X$
\cite{witten,inst}.  In these points the addition of discrete torsion
gives a different component of the moduli space of $\cN=(4,4)$
SCFT. The marginal deformation away from the symmetric product is a
$\Z_2$ twist field. Discrete torsion removes this marginal operator.
The singularities get frozen in.  This component of the $c=6k$
$\cN=(4,4)$ SCFT moduli space is therefore described just by the
moduli of $X$. In particular there is no way to deform the model to a
regime where the weakly coupled supergravity approximation of the dual
formulation as string theory on $AdS_3 \times S^3 \times X$ makes
sense.

One of the striking properties of the relation between space-time
physics and the D1-D5 CFT is that the elusive Ramond-Ramond gauge
fields appear as more traditional $B$-fields in the sigma model. Since
we are discussing here a discrete $B$-field on the symmetric product,
the spacetime interpretation would seem to be some new discrete RR
flux in the $AdS_3 \times S^3 \times X$ string theory. It would be
very interesting to identify directly this RR flux.

\medskip

For orbifold CFT with gauge group $G$ there are natural
interpretations of the cohomology groups $H^i(G,U(1))$ for
$i=1,2,3$. The group $H^1(G)$ labels the one-dimensional
representations and can be used to twist the original $G$ action. For
the permutation group we have $H^1(S_N)=\Z_2$ and this just means that
we can choose the short strings to be either fermionic or bosonic.

We have discussed at length the effect of $H^2(S_N)$, how it effects
the statistics of the long strings. So this leaves the possible
interpretation of the cohomology group $H^3(S_N)$. For a general
orbifold it classifies the possible chiral structures of the
CFT. Alteratively, it gives the possible three-dimensional topological
discrete Chern-Simons gauge theories with gauge group $S_N$
\cite{dw}. It is an interesting fact that for the symmetric group there is a
well-known factor
\be
\Z_{24} \subset H^3(S_N,U(1)).
\ee
As far as I understood, this occurrence of the number $24$ is directly
related to the famous $c/24$ in CFT, the framing ambiguity in
three-manifold invariants, and the Euler number of $K3$. It would be
fascinating to know if it has any application in terms of
second-quantized strings.

\vspace{8mm}

{\noindent \bf Acknowledgements}

\vspace{2mm}

I would like to acknowledge useful discussions with A. Adem, J. de
Boer, F. Cohen, M. Hopkins, T. Mrowka, G. Segal, and C. Vafa.  I also
wish to thank the MIT Mathematics Department, and in particular
I. Singer, for the warm hospitality while this work was done.

\renewcommand{\Large}{\normalsize}
\renewcommand{\tt}{}

 \end{document}